\documentclass[aps,nofootinbib,showpacs,preprint,tightenlines]{revtex4-1}

\usepackage{graphicx}
\usepackage{amsmath}
\usepackage{bm}
\usepackage{color}
\usepackage{amssymb}

\newcommand{\beq}{\begin{equation}}
\newcommand{\eeq}{\end{equation}}
\newcommand{\bqa}{\begin{eqnarray}}
\newcommand{\eqa}{\end{eqnarray}}
\newcommand{\del}{\partial}
\newcommand{\half}{\frac{1}{2}}

\newcommand{\gsim}{\hspace*{0.2em}\raisebox{0.5ex}{$>$}
     \hspace{-0.8em}\raisebox{-0.3em}{$\sim$}\hspace*{0.2em}}
\newcommand{\lsim}{\hspace*{0.2em}\raisebox{0.5ex}{$<$}
     \hspace{-0.8em}\raisebox{-0.3em}{$\sim$}\hspace*{0.2em}}
\def\mqo2{{\!\!\!}}

\begin{document}


\title{On the Efimov Effect in Higher Partial Waves}
\author{K.\ Helfrich}

\author{H.-W.\ Hammer}
\affiliation{Helmholtz-Institut f\"ur Strahlen- und Kernphysik (Theorie)\\
and Bethe Center for Theoretical Physics,
 Universit\"at Bonn, 53115 Bonn, Germany}

\date{\today}

\begin{abstract}
Using the framework of effective field theory,
we present a detailed study of the Efimov effect in higher 
partial waves for systems of two identical particles and a third
distinguishable particle. Depending on the total angular momentum 
$L$, the two identical particles must be bosons or fermions.
We derive analytical expressions for the elastic and inelastic
atom-dimer scattering cross sections as well as the atom-dimer relaxation
rate at the dimer breakup threshold.
For the experimentally most relevant case of P-waves,
we numerically calculate the atom-dimer scattering cross sections and 
relaxation rates as a function of the scattering 
length, three-body parameter, and mass ratio for energies
below breakup threshold.
\end{abstract}
\maketitle

\section{Introduction}
\label{sec:intro}
The last few years saw an enormous progress in the field of Efimov physics, 
theoretically as well as experimentally~\cite{Ferlaino}. 
Already in 1970, Efimov predicted 
the existence of universal three-body bound states with a geometric spectrum
for identical bosons at infinite scattering length $a$~\cite{Efimov70}
\beq
\label{eq:bind}
B_t^{(n)}=\left(e^{-2\pi/s_0}\right)^{n-n_*}\hbar^2\kappa_*^2/m\,,
\eeq
where $m$ is the mass of the particles, $s_0\approx 1.00624$ is a transcendental
number, and $\kappa_*$ is the binding wave number of the state labelled $n_*$.
This leads to an accumulation point at $E=0$ and a discrete scale invariance with
a scaling factor of $e^{\pi/s_0}\approx 22.7$. Away from unitarity, there is only 
a finite number of bound states. 
As the inverse scattering length is varied from negative to positive values, an 
Efimov state appears at the three-particle
scattering threshold at $a=a_-$, crosses the unitary limit at $1/a=0$ with the energy
given in Eq.~(\ref{eq:bind}), and
vanishes at $a=a_*$ through the atom-dimer threshold. 
Whenever an Efimov trimer hits one of these thresholds, recombination rates get enhanced
\cite{EGB-99,BH01,Braaten:2003yc}. 

Measuring this enhancement,
the Efimov effect has by now been observed in bosonic isotopes of 
cesium~\cite{Kraemer:2006,Knoop:2008}, 
potassium~\cite{Zaccanti:2009}, and lithium~\cite{Gross:2009,Hulet:2009}.
The case of fermionic lithium with three different hyperfine states is 
similar to the bosonic case in many aspects. Because of the three different pair scattering lengths
involved, however, the phenomenology is much richer. After the indirect observation via resonant 
enhancement of the recombination rates~\cite{Ottenstein:2008,Huckans:2008}, the direct 
association of Efimov trimers was also achieved~\cite{Lompe:10,Nakajima:11}. 

Mixed systems also allow for the appearance of the Efimov effect as long as
at least two of the scattering lengths are resonant. This is most easily realised
with two different atomic species, one of which has to be bosonic for the S-wave case. 
If the bosonic species is much heavier than the other one, the scaling factor can become 
considerably smaller and thus more favorable
for the experimental investigation~\cite{Efimov73}. This case has been considered in detail
in~\cite{Helf:10}.  The first experimental measurement of the Efimov effect in
a bosonic rubidium-potassium mixture was reported in~\cite{Barontini:2009}.

However, if the heavy species in a heteronuclear mixture is fermionic, the Efimov effect
is only present in a P-wave channel and only if the masses differ by at least a factor of 
13.61~\cite{Efimov73}. This behaviour can be generalized to higher partial waves. The parameters $s_L$
determining the scaling factor for total angular momentum $L$
can be estimated by~\cite{Efimov73}  
\beq
\label{sell}
s_L^2\approx s_0^2-L(L+1)\,,
\eeq
where $s_0$ characterizes the corresponding scaling factor for zero angular momentum.
$s_L^2\geq 0$ must be fulfilled for the occurrence of the Efimov effect (also see more 
detailed discussion in section~\ref{sec:frame}).
The necessary mass ratios for the P-wave case ($L=1$) could be realised for example with 
$^6$Li-$^{87}$Sr, $^{6/7}$Li-$^{137}$Ba, $^{6/7}$Li-$^{167}$Er, or $^{6/7}$Li-$^{171/173}$Yb. 
Higher partial waves would only become accessible if hydrogen or helium atoms can be used.
Dimer-dimer scattering in heteronuclear mixtures showing a P-wave Efimov effect
was investigated in~\cite{Marc:08}. 

The main focus of this paper is also on such mixtures, prepared as atom-dimer
systems. We study in detail how the Efimov effect
in higher partial waves affects observables such as atom-dimer scattering and atom-dimer relaxation. 
We briefly introduce our effective field theory framework 
and derive analytical expressions for the elastic and inelastic
atom-dimer scattering cross sections at the dimer breakup threshold.
For the P-wave case,
we numerically calculate the atom-dimer scattering cross sections and 
relaxation rates as a function of the scattering 
length, three-body parameter, and mass ratio below dimer breakup threshold.
Finally, some quantities for systems without Efimov effect are also computed.

\section{Framework}
\label{sec:frame}
In the following, we investigate various heteronuclear atomic systems in 
detail and closely follow the formalism and conventions of Ref.~\cite{Helf:10}. 
For convenience, we set $\hbar=1$ but restore the dimensions for final
results. We consider systems 
consisting of two different atomic species, where the occurring 
trimers, Efimov or non-Efimov, are built of one atom of type 1 and two atoms 
of type 2. The unlike particles have a resonant S-wave interaction which can be
tuned using a Feshbach resonance whereas the
interaction between identical particles can be neglected.
The corresponding Lagrangian is given by~\cite{Helf:10}
\bqa
\label{eq:Lagrangian}
{\cal L}&=&\Psi^\dagger_1\left(i\del_t+\frac{\nabla^2}{2m_1}\right)\Psi_1
+\Psi^\dagger_2\left(i\del_t+\frac{\nabla^2}{2m_2}\right)\Psi_2\nonumber \\
&+&g_2d^\dagger d-g_2\left(d^\dagger\Psi_1\Psi_2
+\Psi^\dagger_1\Psi^\dagger_2d\right)-\frac{g_3}{4}d^\dagger d\Psi_2^\dagger \Psi_2+\cdots\, ,
\eqa
where $m_{1/2}$ denotes the mass of particles of species $1/2$, $g_{2/3}$
are the bare two-body and three-body coupling constants, 
and $d$ is an auxiliary field for a dimer consisting of particle species 1 and 2. 
The first two terms in Eq.~(\ref{eq:Lagrangian}) are the kinetic terms for species 
1 and 2, while the third term is the kinetic term for the non-dynamical auxiliary dimer field
which mediates the interactions between the particles of species 1 and 2.
It becomes dynamical through the coupling to particle loops. The fourth and fifth
terms generate the S-wave contact interactions between particles of species 1 and 2 and 
a three-body interaction between one particle of species 1 and two particles of species 2.\footnote{To this order, the coefficient of the dimer kinetic term is not independent of the interaction between particles 1 and 2 and can be chosen as $g_2$ for convenience.}
The ellipses stand for higher-order terms containing more fields
and/or derivatives. As in Ref.~\cite{Helf:10}, we denote the ratio
of the masses of particle species 1 and 2 by $\delta\equiv m_1/m_2$.
In this study, however, we explicitly focus on higher partial
waves with total angular momentum $L>0$. 

Because of symmetry reasons,
the Efimov effect can only occur in even angular
momentum channels if the two like particles are bosons and in odd angular momentum 
channels if they are fermions~\cite{Niel:01}. In the following, we will refer to
the first case as {\it bosonic} and the second case as {\it fermionic} for simplicity.
The nature of the third particle is not relevant for our purpose.
For inverse mass ratios $\delta^{-1}$ larger 
than a critical ratio
$\delta^{-1}_{c,L}$, Efimov physics can be observed~\cite{Efimov73}. At  
$\delta^{-1}_{c,L}$ and beyond,
the angular momentum barrier is overcome by the attractive interaction 
between unlike particles  (cf.~Eq.~(\ref{sell})). The light particle 
can be thought of as an exchange particle
between the two heavy atoms. In this case, the ``fall to the center'' 
phenomenon typical for the Efimov effect can happen. 
A (hybrid) Born-Oppenheimer description has been used in the limit of a very
light particle of species 1~\cite{Marc:08,Efre:09}.
In the case of $L=1$, the mass ratio must satisfy 
$\delta^{-1}\gsim 13.61=\delta^{-1}_{c,1}$~\cite{Efimov73, Petrov:03}.
The D-wave Efimov effect starts at 
$\delta^{-1}\gsim 38.63=\delta^{-1}_{c,2}$~\cite{Efimov73,Kartavtsev:07} and its observation
would always be obscured by the already present S-wave effect. 
Consequently, the fermionic P-wave case
is the only relevant one in cold atom experiments
besides the S-wave bosonic case. 
The mass ratios $\delta^{-1}$ for some possible 
mixtures showing the P-wave Efimov effect are given in 
Table~\ref{tab:massratios}. 
\begin{table}
\begin{tabular}{lcccccc}
\hline
\hline
       &\ \ $^{87}$Sr\hspace{.5cm} &\ \  $^{137}$Ba\hspace{.5cm} &\ \  $^{167}$Er\hspace{.5cm} 
       &\ \ $^{171}$Yb\hspace{.5cm} &\ \ $^{173}$Yb\hspace{.5cm} \\
\hline
$^6$Li\ \ & 14.5  & 22.8   &  27.8  & 28.5  & 28.8\\
$^7$Li\ \ & -         & 19.6  & 23.9  & 24.4  & 24.7\\ 
\hline
\hline
\end{tabular}
\caption{Mass ratios $\delta^{-1}$ for possible experimental mixtures 
showing the P-wave Efimov effect.}
\label{tab:massratios}
\end{table}

It is straightforward to derive Feynman rules and to obtain the full 
dimer propagator
from the Lagrangian in Eq.~(\ref{eq:Lagrangian}) (for more details, see 
Refs.~\cite{Braaten:2004rn,Helf:10}). In the three-body system, we obtain
an integral equation for the off-shell atom-dimer scattering amplitude 
${\cal A}_L(p,k;E)$ known as the STM equation~\cite{STM}. 
The amplitude depends on the relative momenta of the atom and the
dimer in the initial state, $k$, and in the final state, $p$, as
well as on the total energy $E$.
All three-body observables can be obtained from 
this amplitude evaluated in appropriate kinematics.
Projecting on total angular momentum $L$, the equation becomes
\bqa
\label{eq:STM}
{\cal A}_L(p,k;E)&=&(\pm 1)\frac{2\pi m_1}{a\mu^2}
\frac{1}{pk}(-1)^L Q_L\left(\frac{p^2+k^2-2\mu
    E}{2pk\mu/m_1}\right)\nonumber \\
&+&(\pm 1)\frac{m_1}{\pi\mu}\int_0^{\Lambda_c} dq\,\frac{q}{p}(-1)^L Q_L
\left(\frac{p^2+q^2-2\mu
    E}{2pq\mu/m_1}\right)\nonumber \\
&\times& \frac{{\cal A}_L(q,k;E)}{-1/a+\sqrt{-2\mu(E-q^2/(2\mu_{AD}))}}\,,
\eqa
where the prefactor $+1$ corresponds to the atoms of species 2 being 
bosons and $-1$ to fermions, respectively.
Moreover, $\mu=m_1m_2/(m_1+m_2)$ is the reduced mass of two unlike atoms, 
$\mu_{AD}=m_2(m_1+m_2)/(m_1+2m_2)$ 
is the reduced mass of an atom and a dimer, 
\beq
Q_L(z)=\half\int_{-1}^1dx\,\frac{P_L(x)}{z-x}
\eeq
is a Legendre function of the second kind, and $P_L(x)$ is a Legendre 
polynomial. 
The log-periodic dependence of the 
three-body interaction $g_3$ on the cutoff has been used to absorb the complex 
three-body parameter into the momentum cutoff 
$\Lambda_c$ in Eq.~(\ref{eq:STM})~\cite{Braaten:2003yc}.
The integration in Eq.~(\ref{eq:STM}) is along a 
straight line in the complex plane from zero to $\Lambda_c$. The physical values 
of the amplitude ${\cal A}$ 
are obtained by evaluating  Eq.~(\ref{eq:STM}) once more for real $p=k$ after 
the solution along the complex contour has been obtained.
The absolute value of the cutoff $\Lambda_c$
is proportional to the binding momentum of the deepest Efimov state,
whereas the complex phase determines the 
width of the Efimov trimers,
$\Lambda_c \propto e^{i\eta_*/s_L} \kappa_*$. Physically, the parameter $\eta_*$
takes into account the effects of deeply-bound dimers which provide
decay channels for the Efimov trimers~\cite{Braaten:2003yc}. 
For S-wave Feshbach resonances, $\eta_*$  is typically of the order 0.1.
In the limit of small $\eta_*/s_L$, the width of Efimov trimers is given by
\beq
\Gamma_t \approx \frac{4 \eta_*}{s_L} \left(
B_t + \frac{\hbar^2}{2\mu a^2}\right)\,,
\eeq
where $B_t$ is the trimer energy (cf. Eq.~(259) in Ref.~\cite{Braaten:2004rn}). Thus for small $\eta_*/s_L$, the width 
becomes small.
Without knowing $\eta_*$, a quantitative prediction is not possible,
but we generally expect lifetimes comparable to the S-wave case if 
$s_L$ is of order one.

In the channels without
the Efimov effect (odd angular momenta for bosons, even angular
momenta for fermions, or for $\delta^{-1}<\delta^{-1}_{c,L}$), 
the absolute value of the momentum cutoff $\Lambda_c$ can be taken to infinity.
The three-body interaction in Eq.~(\ref{eq:Lagrangian})
is higher order in these channels and all 
observables are in leading order determined by the scattering length alone.

\section{Scaling factor and resonance positions}
If more than one Efimov resonance feature can be measured in an experiment,
the scaling factor $\exp(\pi/s_L)$ can be deduced. 
The quantity $s_L$ can be computed analytically
by considering the large momentum behaviour of 
Eq.~(\ref{eq:STM})~\cite{Danilov,Beda:99,Griesshammer:2005ga}.
In this limit the energies and inverse scattering lengths
can be neglected compared to the momenta $p$ and $q$, the inhomogeneous 
term as well as purely polynomial terms in the integral kernel are suppressed, 
and the momentum integration can be extended to
infinity. This leads to the equation
\bqa
\label{eq:STMasy}
\tilde{\cal A}_L(p) 
&=&(\pm 1)(-1)^L\frac{m_1}{\pi\mu} \sqrt{\frac{\mu_{AD}}{\mu}}
\int_0^{\infty} \frac{dq}{q}\,P_L\left(\frac{p^2+q^2}
{2pq\mu/m_1}\right)\,Q_0\left(\frac{p^2+q^2}
{2pq\mu/m_1}\right) \tilde{\cal A}_L(q)\,,
\eqa
where we have defined $\tilde{\cal A}_L(p) \equiv p{\cal A}_L(p,k;E)$.
Since the equation is scale invariant, it has
power law solutions. If the Efimov effect is present, the exponent is
complex: $\tilde{\cal A}_L(p)\propto p^{\pm is_L}$.
Identifying the right hand side of Eq.~(\ref{eq:STMasy}) as a Mellin 
transform, we then obtain a transcendental equation for $s_L$:
\bqa
\label{eq:larbitrary}
1&=&\pm\,\frac{(-1)^L}{\sin(2\phi)}\sum_{k=0}^{k_{max}}
\frac{(2L-2k)!}{(L-k)!\,k!}\,\frac{(-1)^k}{2^{2L-2k}(\sin\phi)^{L-2k}}
\nonumber\\
&&\times\,\sum_{m=0}^{L-2k}\frac{1}{m!\,(L-2k-m)!}\,
\frac{2}{is_L+2m-L+2k}\,\frac{\sin\left[(is_L+2m-L+2k)\phi\right]}
{\cos\left[(is_L+2m-L+2k)\frac{\pi}{2}\right]}\,,
\eqa
where we defined
\beq
\phi=\arcsin\frac{1}{\delta+1}\, ,
\eeq
and
\beq
k_{max}=\left\lbrace
      \begin{array}[c]{ll}
        L/2 & \quad \textnormal{if}\ L\ \textnormal{is even}\\
(L-1)/2& \quad \textnormal{if}\ L\ \textnormal{is odd}
\end{array}
    \right.\,.
\eeq
In the case $L=1$, Eq.~(\ref{eq:larbitrary}) reduces to
\beq
\label{eq:l1}
1=\frac{1}{2\sin^2\phi\cos\phi}\left[\frac{1}{i
    s_1-1}\frac{\sin[(is_1-1)\phi]}{\cos[(is_1-1)\pi/2] }+\frac{1}{i
      s_1+1}\frac{\sin[(is_1+1)\phi]}{\cos[(is_1+1)\pi/2]}\right]\,.
\eeq
The corresponding equation for $L=2$ is given by
\bqa
\label{eq:l2}
1&=&\frac{3}{8\sin^3\phi\cos\phi}\left[\frac{1}{i
    s_2-2}\frac{\sin[(is_2-2)\phi]}{\cos[(is_2-2)\pi/2]}+\frac{1}{i
      s_2+2}\frac{\sin[(is_2+2)\phi]}{\cos[(is_2+2)\pi/2]}\right.\nonumber\\
&&+\left.\frac{2-4/3
        \sin^2\phi}{is_2}\frac{\sin[is_2\phi]}{\cos[is_2\pi/2]}\right]\, .
\eqa

\begin{figure}[htb]
\begin{center}
\includegraphics[clip,width=10cm]{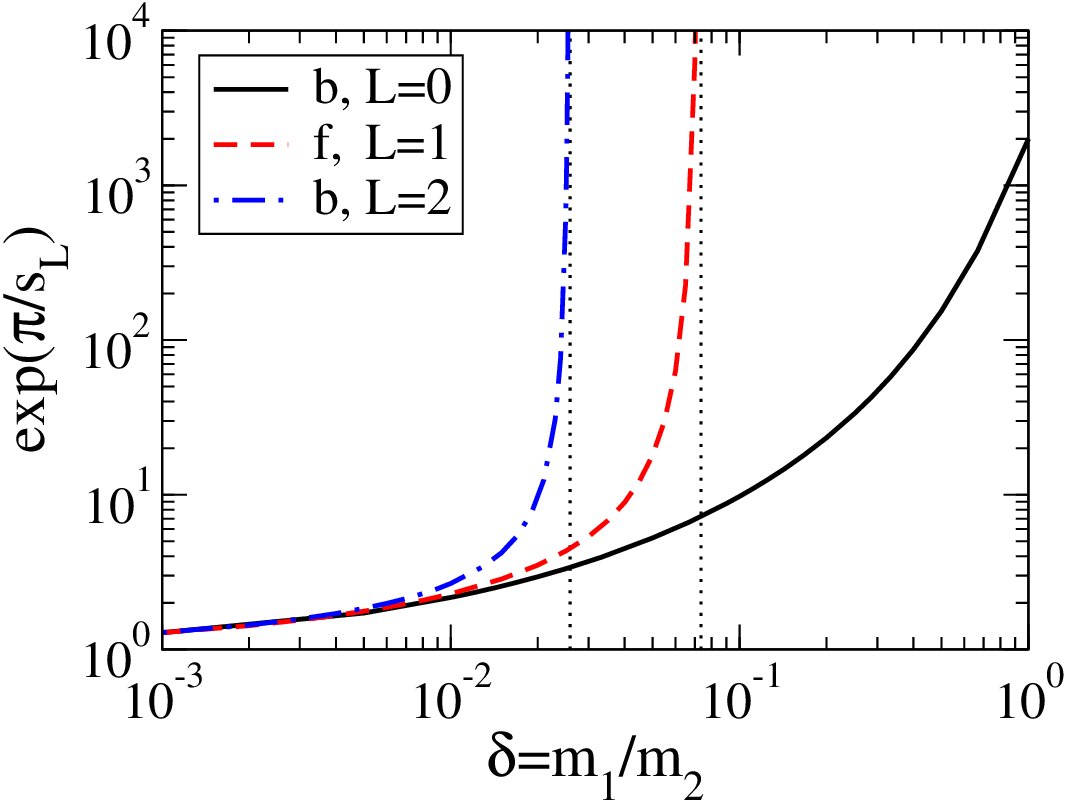}
\caption{Scaling factors $\exp(\pi/s_L)$ for bosons (b) and fermions (f)
 as a function of the mass ratio $\delta=m_1/m_2$ for $L=0,1,2$, respectively.
 The vertical dotted lines indicate the critical mass ratios $\delta_{c,1}$ 
 and $\delta_{c,2}$ for the Efimov effect with $L=1,2$.}
\label{fig:scalefac}
\end{center}
\end{figure}
An equivalent equation for general $L$ 
using hypergeometric functions was derived 
by Nielsen and coworkers~\cite{Niel:01}. The results of 
Eqs.~(\ref{eq:l1}) and (\ref{eq:l2}) coincide with the ones 
obtained by making use of Eq.~(117) in~\cite{Niel:01}. 
The critical mass ratios $\delta_{c,L}$ 
are obtained by observing when $s_L$ tends to zero. We find 
$\delta^{-1}_{c,1}=13.61$ and 
$\delta^{-1}_{c,2}=38.63$ 
in agreement with previous determinations~\cite{Petrov:03, Kartavtsev:07}.
The resulting scaling factors $\exp(\pi/s_L)$ are shown in 
Fig.~\ref{fig:scalefac}
as  a function of the mass ratio $\delta=m_1/m_2$.
The critical mass ratios $\delta_{c,1}$ and $\delta_{c,2}$
for $L=1,2$ can be read off from the positions where the scaling factor
diverges (indicated by the vertical dotted lines).
For $L\to \infty$, the critical mass ratio approaches zero.
For $\delta \to 0$, all scaling factors approach unity corresponding 
to the limit $s_L \to \infty$.

\begin{figure}[ht]
\begin{center}
\includegraphics[clip,width=10cm]{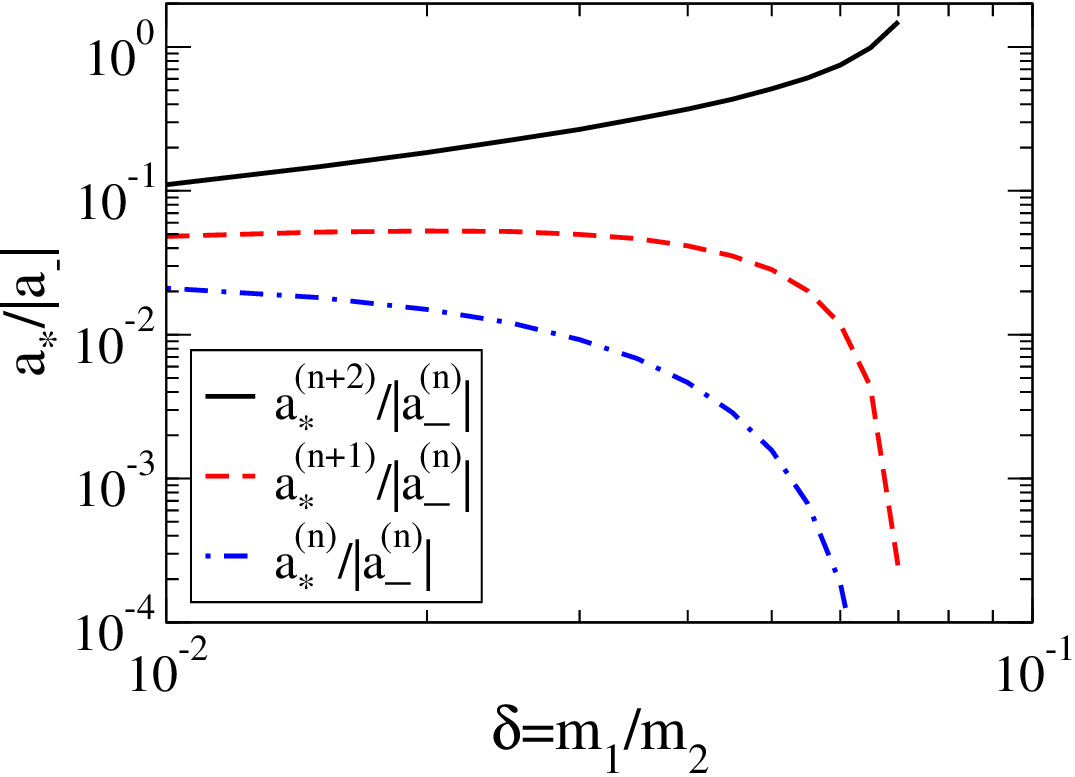}
\caption{The ratio $a_*/|a_-|$  for following one Efimov state and for
  comparing neighbouring states as a function of the mass ratio $\delta=m_1/m_2$ 
  in the case $L=1$.}
\label{fig:ratio}
\end{center}
\end{figure}

Another important observable is the ratio $a_*/|a_-|$ that compares the values of
the scattering length $a_*$ and $a_-$ at which Efimov trimers cross the atom-dimer and three-particle
thresholds, respectively. This ratio can be measured 
experimentally if at least one resonance feature 
is seen for negative scattering length and one in the atom-dimer system. 
The calculated ratios for $L=1$ are shown in Fig.~\ref{fig:ratio}
for following one Efimov state and for
comparing neighbouring states as a function of the mass ratio $\delta=m_1/m_2$.

\section{Analytical results for atom-dimer scattering}
The scattering of atoms and dimers can be directly related to
the STM equation~(\ref{eq:STM}) with equal incoming and outgoing
momenta. The scattering amplitude is given by
\beq
\label{eq:scattamp}
f_L(k)=\frac{k^{2L}}{k^{2L+1}\cot\delta_{AD,L}(k)-ik^{2L+1}}
=\frac{\mu_{AD}}{2\pi}{\cal A}_L(k,k;E)\,,
\eeq
where the energy $E$ and center-of-mass momentum $k$ are related 
by $E=k^2/(2\mu_{AD})-B_d$. 
The dimers are built of two unlike particles and their binding 
energy is given by $B_d=1/(2\mu a^2)\,$.
The dimer breakup threshold  at $E=0$ corresponds to $k=\sqrt{\mu_{AD}/\mu}\,/a\equiv k_{D}$.
It is useful to define an energy dependent {\it scattering length},
\beq
\label{eq:atildeAD}
\tilde{a}_{AD,L}(k)=\frac{-1}{k^{2L+1}\cot\delta_{AD,L}(k)}\,.
\eeq
Note that this quantity does not have the dimension of a length for 
$L>0$. We can now calculate the elastic atom-dimer scattering 
cross section in the $L$th partial wave
\beq
\label{eq:s_elADnum}
\sigma_{AD,L}^{\rm (el)}(k)=(2L+1)4\pi\left|f_L(k)\right|^2\,.
\eeq
The total cross section can be obtained with the help of the optical theorem
\beq
\label{eq:s_totADnum}
\sigma_{AD,L}^{\rm (tot)}(k)=(2L+1)\frac{4\pi}{k}
\textnormal{Im}f_L(k)\,,
\eeq
and the inelastic cross section $\sigma_{AD,L}^{\rm (inel)}(k)$ 
by subtracting Eqs.~(\ref{eq:s_totADnum}) and
(\ref{eq:s_elADnum}).

At $E=0$, it is also possible to deduce these quantities analytically 
using the methods of Section IV of Ref.~\cite{Helf:10}. 
If the Efimov effect is present, the S-matrix element for elastic
atom-dimer scattering can be written as
\beq
S_L=-e^{2i\sigma_L}\cosh(\pi s_L+is_L \log(a/a_{0*})-\eta_*)/
\cosh(\pi s_L-is_L \log(a/a_{0*})+\eta_*)\,,
\label{eq:SL}
\eeq
where $\sigma_L$ is a real number and $a_{0*}$ determines the position 
of the minima in the three-body 
recombination rate for positive scattering length
and in the elastic atom-dimer cross section.
Hence, the elastic cross section can be expressed as
\bqa
\label{eq:s_elADana}
\hspace{-.3cm}\sigma_{AD,L}^{\rm (el)}(E=0)&=&(2L+1)\frac{\pi}{k_{D}^2}\left|S_L-1\right|^2\nonumber\\
&=&(2L+1)\,4\pi a^2\,\frac{\delta(\delta+2)}{(\delta+1)^2}
\frac{\sinh^2(\pi s_L) \left\{\sinh^2(\eta_*)+\sin^2[s_L \log(a/a_{0*})]
\right\}}{\sinh^2(\pi s_L+\eta_*)
+\cos^2[s_L \log(a/a_{0*})]}\,,
\eqa
where $k_{D}a =(\delta+1)/\sqrt{\delta(\delta+2)}\,$
was used.
The inelastic cross section is given by
\bqa
\label{eq:s_inelADana}
\sigma_{AD,L}^{\rm (inel)}(E=0)&=&(2L+1)\frac{\pi}{k_{D}^2}\left(1-\left|S_L\right|^2\right)\nonumber\\
&=&(2L+1)\,\pi a^2\,\frac{\delta(\delta+2)}{(\delta+1)^2}
\frac{\sinh(2\pi s_L)\sinh(2\eta_*)}{\sinh^2(\pi s_L+\eta_*)+
\cos^2[s_L \log(a/a_{0*})]}\,.
\eqa

Atom-dimer relaxation is the process where an atom and a shallow dimer
collide and an energetic deep dimer and atom are ejected. It is one of the 
main loss processes in mixtures of atoms and dimers.
The atom-dimer relaxation rate constant $\beta$ is defined by the rate equation
\beq
\frac{d}{dt}n_D=\frac{d}{dt}n_A=-\beta n_Dn_A\,,
\eeq
where $n_{A/D}$ denotes the number densities of atoms and dimers, respectively.
The relaxation rate is directly related to the inelastic scattering 
cross section,
\beq
\beta_L(E)=\frac{k}{\mu_{AD}}\sigma^{{\rm (inel)}}_{AD,L}(E)\,.
\label{eq:betaL}
\eeq
At $E=0$, we therefore obtain
\beq
\beta_L(E=0)=(2L+1)\pi\frac{\sqrt{\delta(\delta+2)}^3}{(\delta+1)^2}\,\frac{\sinh(2\pi s_L)\sinh(2\eta_*)}{\sinh^2(\pi s_L+\eta_*)+\cos^2[s_L\log(a/a_{0*})]}\,\frac{\hbar a}{m_1}\,.
\eeq
We note that our analytical results for the inelastic
atom-dimer rate at zero energy are in
agreement with the general scaling law derived 
by D'Incao and Esry~\cite{dIncao06}.
At the atom-dimer threshold,  $E=-B_d$, only the S-wave contribution
survives. This case was studied in detail in Ref.~\cite{Helf:10}.
For all other angular momenta $L>0$, the atom-dimer scattering 
amplitude, Eq.~(\ref{eq:scattamp}), vanishes for $k\rightarrow 0$.

Another important process in cold atom experiments is three-body recombination.
This process can happen in a mixture if not all atoms of species 1 are bound in
dimers. The atom of species 1 and an atom of species 2 form a dimer, shallow
or deep, and 
another atom of species 2 balances energy and momentum. Typically, all 
three atoms are lost if this process occurs in a trap. For angular 
momenta $L>0$, however, the three-body recombination rate vanishes at $E=0$.

In the following, we present numerical results for the atom-dimer
observables discussed above
focussing on the experimentally most relevant case $L=1$.

\section{Numerical results}

\subsection{Atom-dimer observables without Efimov effect}
\begin{figure}[ht]
\begin{center}
\includegraphics[clip,width=10cm]{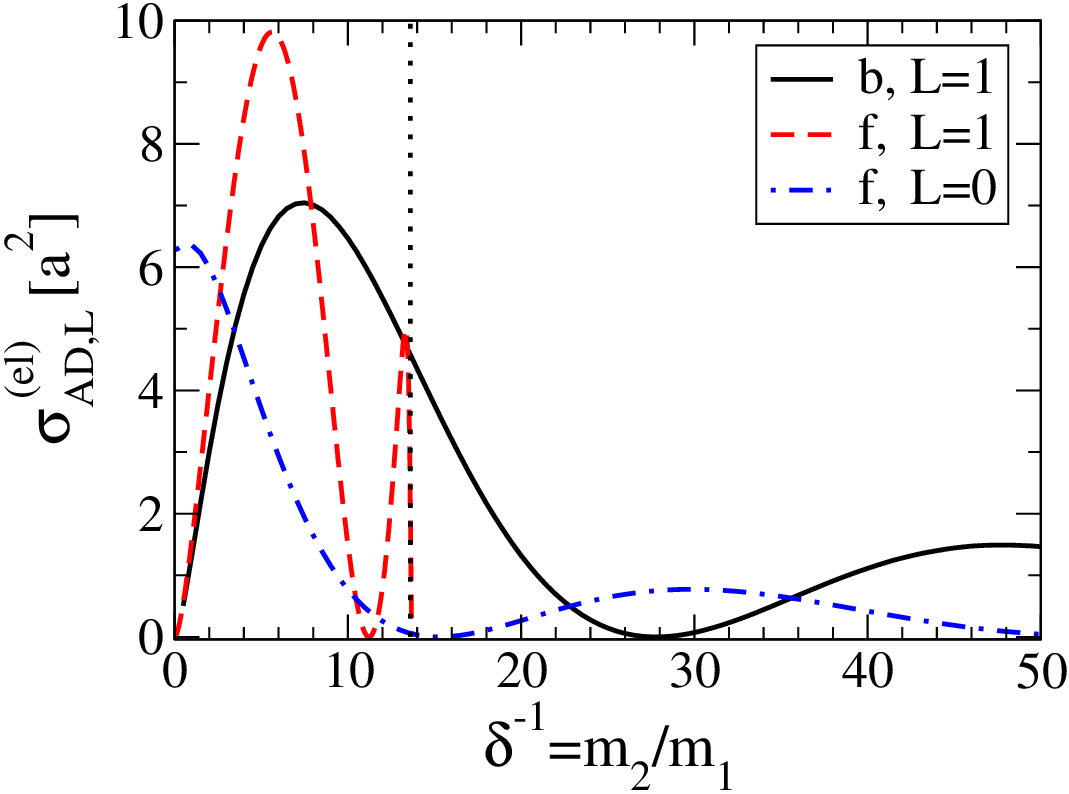}
\caption{The elastic atom-dimer scattering cross sections for bosons (b) and 
fermions (f) in the P-wave channel and for fermions in the S-wave channel at 
$E=0$ as a function of the mass ratio $\delta^{-1}$. The vertical dotted
line indicates the critical mass ratio for fermions in the P-wave channel,
$\delta^{-1}_{c,1}=13.61$.}
\label{fig:crosssec}
\end{center}
\end{figure}

In this subsection, we only consider the case $\eta_* =0$ which corresponds to 
no deeply-bound dimers. As a consequence, the inelastic cross section
vanishes below the dimer breakup threshold.
In the P-wave channel, there is no Efimov effect for bosons.
The total (=elastic) atom-dimer scattering cross section for $E=0$ 
(at dimer breakup)  is shown as the solid line in 
Fig.~\ref{fig:crosssec}. Note that for increasing $\delta^{-1}$,
the cross section does not tend monotonically to zero 
but rather oscillates with diminishing amplitude. 
This oscillation is not due to the crossing of bound three-body states
with the atom-dimer threshold
as three-body bound states are not present in this
system (see below for the fermionic case). 
Using Eq.~(\ref{eq:s_elADnum}),
the elastic cross section at threshold can be written as
\beq
\label{eq:ADcrossth}
\sigma_{AD,L}^{\rm (el)}(k_D)=(2L+1)4\pi a^2\,\frac{(1+2\delta^{-1})}{
(1+\delta^{-1})^2} \sin^2\delta_{AD,L}(k_D)\,.
\eeq
The observed oscillation then implies a monotonic dependence
of the elastic phase shift at threshold on $\delta^{-1}$~\cite{Karta:07}
with $\delta_{AD,L}(k_D) =0$ for  $\delta^{-1}=0$.
For decreasing energy, the amplitude of the cross section gets larger and 
the peaks move to larger mass ratios  $\delta^{-1}$. 

A similar behavior is observed for fermions in the S-wave channel.
The atom-dimer scattering length can be determined according to the formula
$a_{AD,0}=-{\cal A}_0(0,0;-B_d)\mu_{AD}/(2\pi)\,$.
This reproduces the results for the mass dependence found by 
Petrov~\cite{Petrov:03} that were confirmed in Refs.~\cite{Iskin:08,Iskin:10}.
The corresponding total cross section is shown in Fig.~\ref{fig:crosssec} 
as dash-dotted line. Again, the oscillation is
not due to bound states and Eq.~(\ref{eq:ADcrossth}) implies a 
monotonic dependence of the threshold phase shift on $\delta^{-1}$.
However, in this case $\delta_{AD,L}(k_D)$ approaches a value slightly
below $\pi/2$ for $\delta^{-1}=0$.
As in the case of bosons with $L=1$,
we find that the amplitude of the cross section gets larger 
for decreasing energy and the peaks 
move to larger mass ratios $\delta^{-1}$. 

For fermions in the  P-wave channel, the Efimov effect only comes into play for 
mass ratios $\delta^{-1}\gsim 13.61$. 
For the region without the Efimov effect, many observables have already been 
calculated~\cite{Petrov:04,Petrov:05,Karta:07,Lev:09,Lev:11,Endo:11}. 
Kartavtsev and Malykh found one three-body bound state for the 
range $8.17260<\delta^{-1}<12.91743$ and 
two three-body bound states for 
$12.91743<\delta^{-1}<13.6069657$~\cite{Karta:07}. They call these states 
{\it universal}, as their binding energies only depend on the dimer binding 
energy, or equivalently, 
the scattering length. The occurrence of these states was 
recently confirmed by Endo et al.~\cite{Endo:11}. They 
also demonstrated the divergence of the atom-dimer scattering length at the 
mass ratios where the universal trimer states appear and how similar behaviour 
occurs for higher angular momenta. We have confirmed these results. 
Kartavtsev and Malykh also found that close to the 
critical mass ratio, the energies of the universal states follow a square-root 
dependence~\cite{Karta:07},
$E-E_c\propto \sqrt{\delta_{c,1}^{-1}-\delta^{-1}}\,$.
An investigation of the behavior of Efimov states for $\delta^{-1}$
slightly above $\delta_{c,1}^{-1}$ would be interesting. The behavior of the 
energies must be non-analytic in $\delta^{-1}-\delta_{c,1}^{-1}$, since
Efimov states can be shifted to any desired energy by adjusting the 
three-body parameter. However, such a study is beyond the scope of the 
present investigation as our numerical calculations converge only
slowly close to the critical mass ratio.

In Ref.~\cite{Karta:07}, atom-dimer elastic scattering at the dimer 
breakup threshold was also calculated. We show this process 
as the dashed line  in Fig.~\ref{fig:crosssec} 
in comparison to the results for 
bosons with $L=1$ and for fermions with $L=0$.
\begin{figure}[ht]
\begin{center}
\includegraphics[clip,width=10cm]{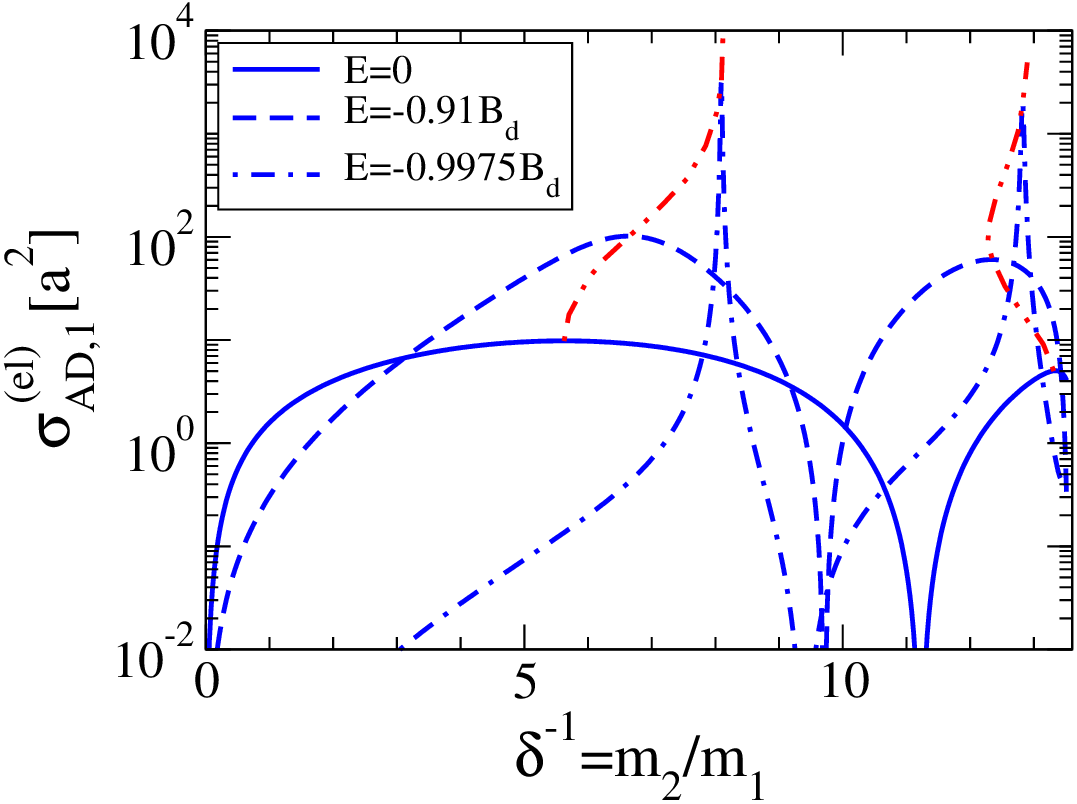}
\caption{The elastic cross section $\sigma^{\rm (el)}_{AD,1}$ 
in units of $a^2$ 
versus the mass ratio $\delta^{-1}$ at the energies $E=0,-0.91B_d,\ 
{\rm and}\ -0.9975B_d$ (solid, dashed, dash-dotted lines). The 
dash-double-dotted lines indicate the position of the two peaks for 
varying energy. 
}
\label{fig:sigferm}
\end{center}
\end{figure}
In order to elucidate the physics of the two peaks in the 
elastic cross section,  we show our results for
$\sigma^{\rm (el)}_{AD,1}$ in Fig.~\ref{fig:sigferm}
as a function of $\delta^{-1}$ for three energies,
$E=0,-0.91B_d,\  {\rm and}\ -0.9975B_d$, i.e.,
at and below the dimer breakup threshold.
The dash-double-dotted lines show how the peak positions move with varying 
energy. While the first peak moves monotonically to larger values of
$\delta^{-1}$ as the energy is decreased, the second peak shows a more 
complicated behavior. Initially, it moves to smaller values
and reaches a minimum $\delta_{\rm min}^{-1} \approx 12.3$ for
$E/B_d\approx -0.938$, before it moves back to larger values of $\delta^{-1}$
as the atom-dimer threshold is approached.
Our results demonstrate that the two-peak structure is 
indeed due to the presence of the two universal three-body bound states. 
The positions of the two peaks move from $\delta^{-1}=5.63$ and 
$13.31$ at $E=0$ to two sharp, $\delta$-function like peaks at 
$\delta^{-1}=8.17$ and $12.9$ for $E=-0.9999B_d$, which are 
the critical values for the occurrence of the three-body bound states. 

\subsection{Atom-dimer observables with Efimov effect}
\begin{figure}[ht]
\begin{center}
\includegraphics[clip,width=10cm]{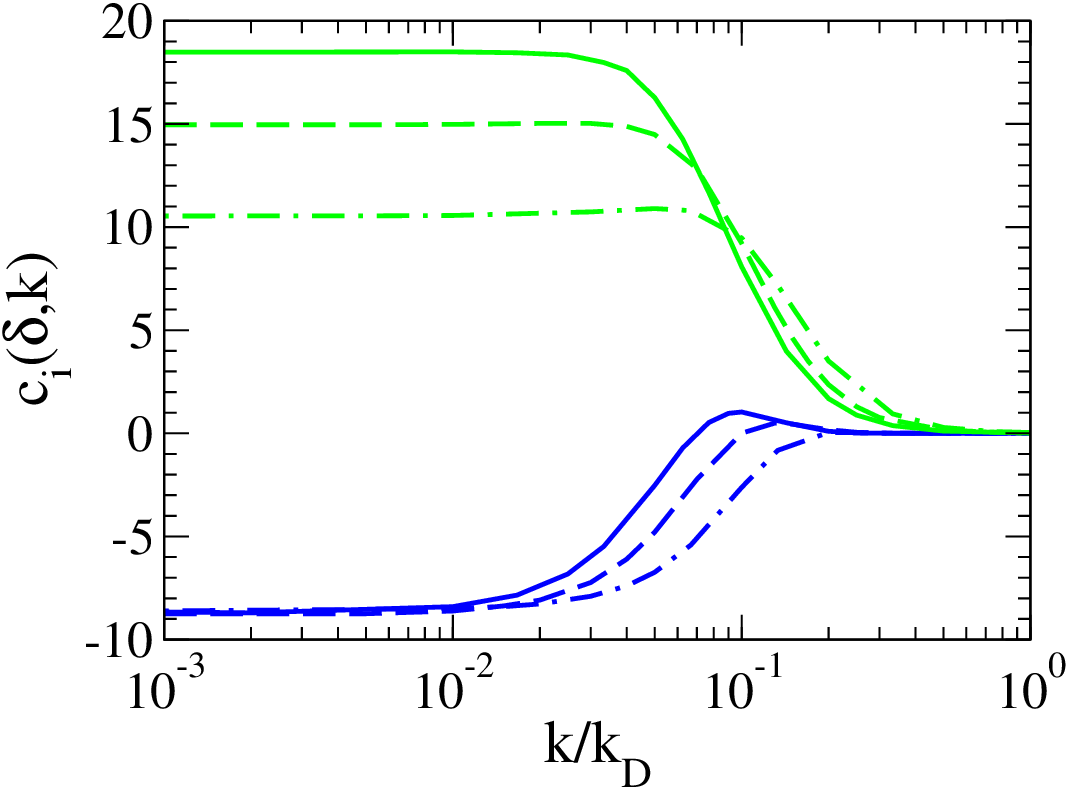}
\caption{Coefficients $c_{1}(\delta,k)$ (dark/blue) and $c_{2}(\delta,k)$
(light/green) 
as a function of $k/k_D$ for 
mass ratios $\delta=0.03,\,0.04,\ {\rm and}\ 0.06$ 
indicated by the solid, dashed, and dash-dotted lines, respectively.}
\label{fig:coeffaAD}
\end{center}
\end{figure}

In the presence of the Efimov effect, the observables do not only depend on 
the mass ratio and 
the energy but they also depend log-periodically on the scattering length. 
From now on, we focus on the case of fermions with $L=1$. We omit
the additional subscript 1 indicating the P-wave channel for notational simplicity.
We find the energy-dependent atom-dimer scattering length for 
$\delta^{-1}>13.61$ calculated 
with Eq.~(\ref{eq:atildeAD}) to be very well approximated by the formula
\beq
\label{eq:atildeADfit}
\tilde{a}_{AD}(\delta,k,a)\equiv\frac{-1}{k^3\cot\delta_{AD}(\delta,k,a)}
=\Bigl\{c_1(\delta,k)+c_2(\delta,k)\cot\bigl[s_1\log(a/a_{*})+i\eta_*\bigr]\Bigr\}a^3\,.
\eeq
We show $c_{1/2}(\delta,k)$ in Fig.~\ref{fig:coeffaAD} for
the mass ratios 
$\delta=0.03,0.04,\ {\rm and}\ 0.06$ as functions of the momentum in units of 
the breakup momentum $k_{D}$ from $0.001\,k_D$ up to the dimer breakup 
threshold.
Interestingly, for all considered mass ratios, $c_1(\delta,k)= -8.7\pm 0.2$ 
for $k\rightarrow 0$. The coefficient $c_2(\delta,k)$ also approaches constant 
values in this limit but this limit depends on the mass ratio. The smaller 
$\delta$, the larger this approached value.
\\
At the dimer breakup threshold, we find
\beq
c_2(\delta,k_{D})\approx k_{D}^{-3}\qquad\mbox{and}\qquad
c_1(\delta,k_{D})\approx 0
\label{eq:appcoeff}
\eeq
to be very good approximations
for $\delta\lsim 0.06$. This behavior is similar to the case of spinless 
bosons~\cite{Braaten:2004rn}. 
From Eq.~(\ref{eq:SL}), we can deduce the atom-dimer
scattering phase shift $\delta_{AD}$ at $E=0$. 
For $\exp(2\pi s_1) \gg \exp(\pm 2\eta_*)$, the expression simplifies
to
\beq
\delta_{AD} = \sigma_1 +s_1 \ln(a/a_{0*}) +i\eta_*\,,
\eeq
and the constraints (\ref{eq:appcoeff}) follow from Eq.~(\ref{eq:atildeADfit}).
Since $s_1$ approaches $0$ as $\delta\to\delta_{c,1}=0.07349$ from below, this
approximation is invalid at larger mass ratios. In this case, we find that
$c_1(\delta,k_{D})$ tends to 
slightly larger and $c_2(\delta,k_{D})$ to slightly smaller  values.

The elastic and inelastic atom-dimer scattering cross sections 
show the typical log-periodic 
dependence on the scattering length. 
For general momenta, they can be approximated with 
Eqs.~(\ref{eq:scattamp})--(\ref{eq:s_totADnum}, \ref{eq:atildeADfit})
and the appropriate coefficients from Fig.~\ref{fig:coeffaAD}. 
At $E=0$, we can compare our numerical calculation 
to the analytical formulae in Eqs.~(\ref{eq:s_elADana}) 
and~(\ref{eq:s_inelADana}). For $\sigma_{AD}^{\rm (inel)}(k_{D})$ which shows only
a weak dependence on $a$, we find generally good agreement. 
For the elastic cross section $\sigma_{AD}^{\rm (el)}(k_{D})$, 
we find very good agreement in the region $\delta\lsim 0.06$.
For larger $\delta$, the numerical calculation becomes
difficult because of the large scaling factor. 

\begin{figure}[ht]
\begin{center}
\includegraphics[clip,width=10cm]{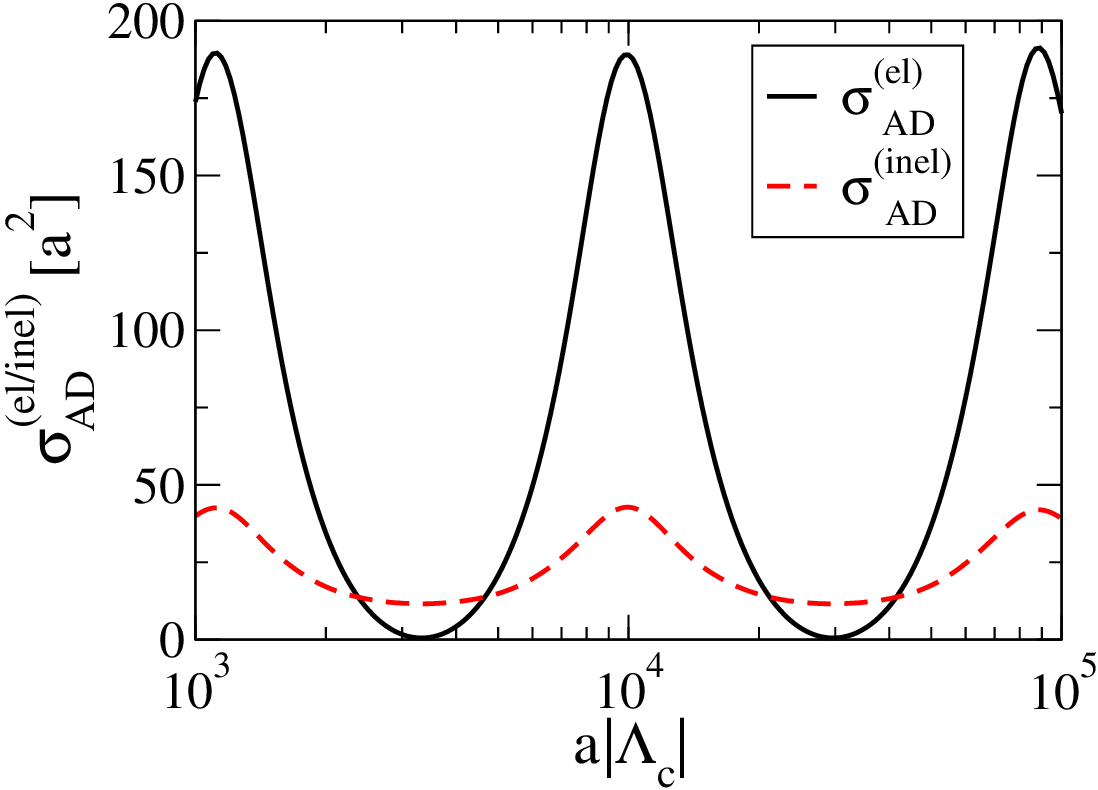}
\caption{Elastic (solid line) and inelastic (dashed line) 
atom-dimer scattering cross sections in units of $a^2$ for 
$\delta=0.04$, $\eta_*=0.1$, and $E=-0.99B_d$ 
versus $a|\Lambda_c|$.}
\label{fig:sigex}
\end{center}
\end{figure}
\begin{figure}[ht]
\begin{center}
\includegraphics[clip,width=10cm]{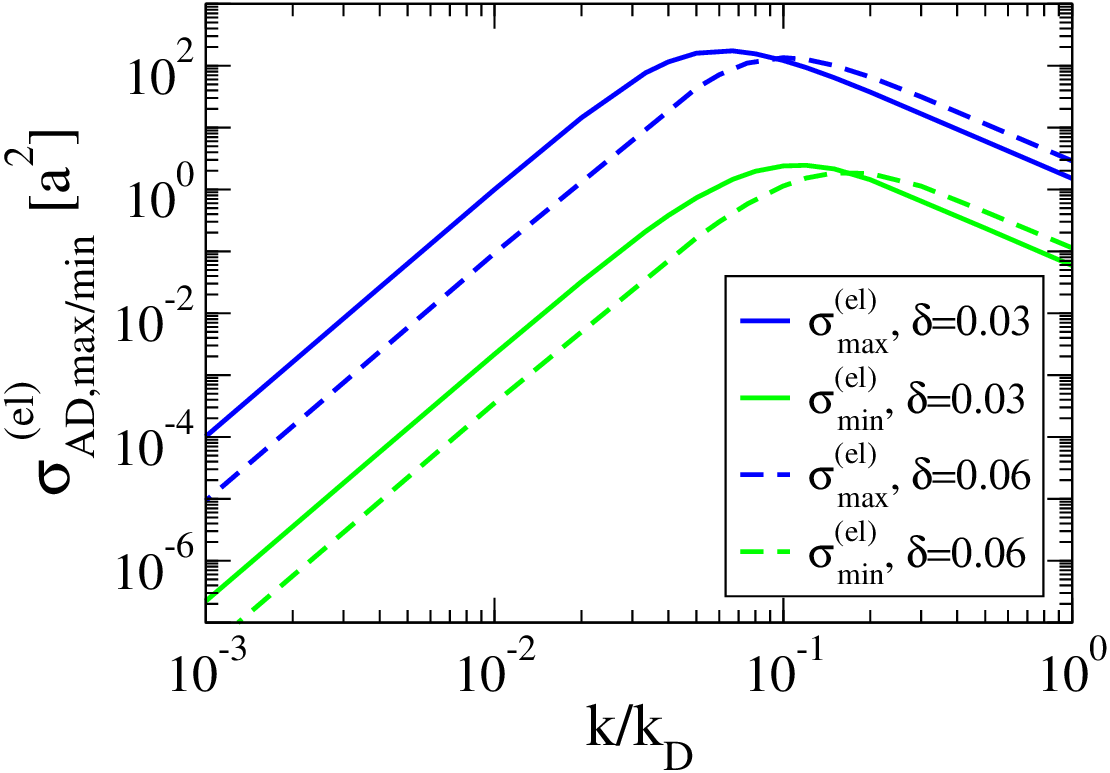}
\caption{Maximal and minimal values of $\sigma_{AD}^{\rm (el)}(k)$ as
a function of $k/k_D$ for $\delta=0.03$ (solid line), $\delta=0.06$ 
(dashed line) and $\eta_*=0.2$.}
\label{fig:sigelmax}
\end{center}
\end{figure}

As an example, we show the elastic and inelastic
cross sections for $\delta=0.04$, $\eta_*=0.1$, 
and $E=-0.99B_d$ in Fig.~\ref{fig:sigex}. This mass ratio 
$\delta$ roughly corresponds 
to the mixtures $^7$Li-$^{171/173}$Yb. The cross sections show
the typical log-periodic dependence on the scattering length.
The values of the maximal and minimal cross section depend 
strongly on the energy and vary over several orders of magnitude.
To demonstrate this dependence, we show 
the maximal and minimal values of $\sigma^{{\rm (el)}}_{AD}$ 
as a function of the center-of-mass momentum $k/k_D$ for 
$\delta=0.03,\, 0.06$ and $\eta_*=0.2$ in Fig.~\ref{fig:sigelmax}. 

\begin{figure}[ht]
\begin{center}
\includegraphics[clip,width=10cm]{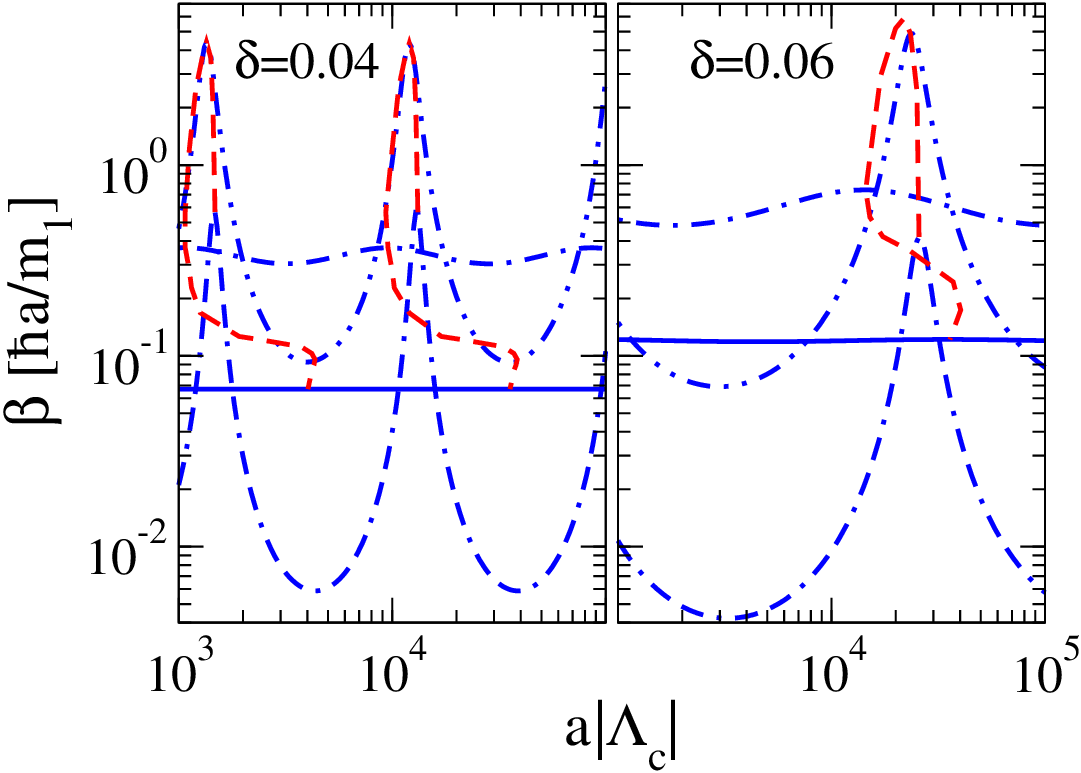}
\caption{Dimer relaxation rate $\beta$ for $\delta=0.04$ and $0.06$ in units of $\hbar a/m_1$ 
 for $\eta_*=0.1$ versus $a|\Lambda_c|$. The double-dash-dotted,
dash-double-dotted, dash-dotted, and solid lines correspond to the energies
$E=-0.9999B_d,\,-0.9984B_d,\,-0.96B_d,\ {\rm and}\ 0$, respectively.
The dashed lines show how the peak position shifts with the energy.}
\label{fig:beta}
\end{center}
\end{figure}

The atom-dimer relaxation rate $\beta$
which can be measured in cold atom experiments is determined 
by the inelastic cross section via Eq.~(\ref{eq:betaL}).
In the case of P-waves, $\beta$ vanishes at the atom-dimer
threshold $E=-B_d$.
In Fig.~\ref{fig:beta}, we show $\beta$ for $\delta=0.04$ and $0.06$,
$\eta_*=0.1$, above threshold for $E=-0.9999B_d,-0.9984B_d,-0.96B_d,\ {\rm and}\ 0$. 
We also give the positions of the peak in $\beta$ for varying energy.
The relaxation rate $\beta$ shows a strong energy dependence as well. Starting from
being zero at the atom-dimer threshold, it develops resonant log-periodic structures for 
larger energies which become less pronounced as the dimer breakup threshold is
approached. The peak position is also strongly energy-dependent and varies by a
factor of four for $\delta=0.04$ and by a factor of three for $\delta=0.06$.
For other mass ratios the qualitative behavior is very similar.

\section{Summary and Outlook}
\label{sec:summ}
In this paper, we have investigated the Efimov effect for heteronuclear systems 
of two identical particles and a third distinguishable particle in higher 
partial waves. The unlike particles were assumed to have resonant S-wave
interactions while the interaction between like particles was
neglected. For even (odd) angular momentum $L$, the two identical 
particles must be bosons (fermions) for the Efimov effect to 
occur~\cite{Efimov73}.
Using an effective field theory framework, we have derived a generalized 
STM equation which describes the off-shell atom-dimer scattering 
amplitude in the total angular momentum channel $L$. All three-body
observables can be extracted from this amplitude taken in appropriate 
kinematics.

We have derived a transcendental equation for the preferred scaling
factor $\exp(\pi/s_L)$ for arbitrary $L$ as a function of the mass ratio 
$\delta$. The numerical results agree well with a 
previously derived equation using hypergeometric functions derived 
by Nielsen and coworkers~\cite{Niel:01}. For the experimentally most relevant
case of the P-wave Efimov effect, we have predicted the ratio of the 
scattering lengths where Efimov states cross the atom-dimer and
three-atom thresholds, $a_*/|a_-|$. This ratio is independent of
the three-body parameter and can be measured in experiment. For the
S-wave case, Barontini et al.~\cite{Barontini:2009} have measured 
the value $a_*/|a_-|=2.7$ in a K-Rb mixture. The universal prediction
for this system is  $a_*/|a_-|=0.52$. 
The Efimov features in the experiment by Barontini et al.\ are not 
deeply in the universal region and thus finite range corrections are 
likely important for this experiment. The leading finite range correction
is given by the  effective range correction, but for further improvement
the van der Waals tail of the interaction has to be included explicitly.
Moreover, there is a finite energy shift of the atom-dimer rescattering resonance 
used in Ref.~\cite{Barontini:2009} to extract $a_*$ that needs 
to be taken into account \cite{Helf:10}.
For P-waves, no experiment has been carried out to date.

The measurement of atom loss rates has played a key role for the observation of
the S-wave Efimov effect in cold atoms~\cite{Ferlaino} and the P-wave
Efimov effect in a Bose-Fermi mixture could be detected in an
analogous  way.  We have derived analytical 
expressions for the elastic and inelastic atom-dimer cross sections
as well as the atom-dimer relaxation rate for arbitrary angular momentum $L$
at the dimer breakup threshold.
For energies below this threshold, we have laid out a framework to
calculate these quantities numerically. 

Using this framework, we have explicitly calculated the atom-dimer 
scattering cross sections for $B_d<E\leq 0$ in
low angular momentum channels without 
the Efimov effect, i.e., bosons in the P-wave and fermions in the S-Wave
channel. Furthermore, we have calculated the cross section 
for fermions in the P-wave channel below the 
critical mass ratio $\delta_{c,1}^{-1}$. The cross section shows two peaks
due to the appearance of two non-Efimov three-body bound states~\cite{Karta:07}.
We have calculated the position of these peaks as a function of the collision 
energy $E$. 

Focussing on the P-wave fermionic channel above the critical mass
ratio, we have  numerically calculated the atom-dimer cross section up to
the dimer breakup threshold.  The cross sections show
the typical log-periodic dependence on the scattering length.
The maximal and minimal cross section values depend 
strongly on the energy and vary over several orders of magnitude.
At the atom-dimer threshold, we found
good agreement with our analytical results.
The atom-dimer cross section below the dimer breakup threshold
can be parametrized by two universal functions  $c_{1}(\delta,k)$  
and $c_{2}(\delta,k)$. We have calculated these functions for several
mass ratios and derived simple analytical expressions for their
values at the  dimer breakup threshold if
$\exp(2\pi s_1) \gg \exp(\pm 2\eta_*)$ is satisfied.
Finally, we have numerically calculated the atom-dimer relaxation rate $\beta$
as a function of the three-body parameter, mass ratio and energy.
As for the bosonic case~\cite{Helf:10}, the position of the relaxation maxima is strongly energy dependent
and not a monotonic function of energy.

In summary, our calculation provides a basis for interpreting
experimental results on the Efimov effect in higher partial waves.
Due to the fermionic nature of the dimers, the preparation of 
the required atom-dimer mixture for the P-wave case should be feasible.
A few experimental groups already study heteronuclear 
mixtures of interest to this work, e.g.\ various Yb-Li mixtures in the 
groups of Takahashi~\cite{Hara:11} 
and Gupta~\cite{Hansen:11}. Other groups investigate heavy species
that could be mixed with Li, e.g.\ fermionic Sr in Grimm's 
group~\cite{Tey:10}. 
A natural extension of our
work would be to calculate the three-body recombination
rate for energies away from the dimer breakup threshold. This could
in principle be done using the methods of Refs.~\cite{BHKP:08,KH:09}.

\acknowledgments
We thank V.\ Efimov for directing our interest towards this subject. 
K.H.\ was supported by the \lq\lq Studien\-stiftung des
Deutschen Volkes'' and by the
Bonn-Cologne Graduate School of Physics and Astronomy.
H.W.H.\ acknowledges support from the the Bundesministerium f\"ur Bildung und 
Forschung, BMBF under Contract No.~06BN9006.


\end{document}